\documentclass[11pt,a4paper]{article}
\pdfoutput=1
\usepackage{jheppub}
\usepackage{subfig}

\newcommand{\del}{\ensuremath{\partial}}
\newcommand{\ba}{\begin{eqnarray}}
\newcommand{\ea}{\end{eqnarray}}

\title{The covariance of multi-field perturbations, pseudo-susy and f$_{\rm NL}$.}

\author[a]{Paul M. Saffin,}

\affiliation[a]{School of Physics and Astronomy, University Park, University of Nottingham,\\ Nottingham NG7 2RD, UK}

\emailAdd{paul.saffin@nottingham.ac.uk}

\abstract{We reconsider cosmological perturbation theory for multi-component scalars, enforcing covariance in field-space, and ensuring that phyical observations are independent of field re-definitions. We use the formalism to clarify some issues in the literature, and use pseudo-supersymmetry to derive various quantities without relying on the slow-roll approximation.}


\begin{document}

\maketitle

\section{Introduction}
Expressing theories in a manner that is manifestly covariant under spacetime co-ordinate transformations ensures a degree of robustness; physical results will not depend upon the users choice of co-ordinates, which is generally regarded as a good thing. However, the spacetime that a field theory lives on is not necessarily the only manifold in a model, if one has $n_\varphi$ scalar fields then we must also consider the manifold parametrized by the scalars, and ensure that our results do not depend upon our choice of parametrization. For example, if we have a complex field $\varphi$ we should be able to use $\varphi=u+iv$ or $\varphi=\rho e^{i\theta}$ as parametrizations without altering physical results. This is something that is well accepted in many areas of particle theory, but has not penetrated all areas of cosmology, with many papers on perturbation theory quoting "general" formulae in a manner that depends upon the choice of field parametrization. For some expressions, such as derivatives of the amount of expansion $N$\cite{Lyth:2005fi},
\ba
\label{eq:2ndDerivN}
\sum_{\alpha,\beta}\frac{\del^2N}{\del\varphi^\alpha\del\varphi^\beta}\frac{\del N}{\del\varphi^\alpha}\frac{\del N}{\del\varphi^\beta},
\ea
it is fairly straightforward to guess what the covariant expression is - simply replace partial with covariant derivatives. However, even here we have to take care of how the expression is derived, as this follows from a Taylor expansion of the local number of e-folds - which {\it does} involve partial rather than covariant derivatives. Other expressions such as the correlation functions of the perturbations $\delta\varphi^\alpha$ of the scalar fields  \cite{astro-ph/0603799,Yokoyama:2008by,astro-ph/0009131,astro-ph/0506056}
\ba
\langle\delta\varphi^\alpha\delta\varphi^\beta\rangle\sim G^{\alpha\beta},
\ea
where $G_{\alpha\beta}$ are the scalar-manifold metric components, are more subtle as the left and right hand sides transform differently under field redefinitions. Such complications are necessary in models where the field-metric is not flat, such as typical supergravity models with non-canonical kinetic terms, but it does not end there. Simply saying that a particular calculation is for flat field-space may not be enough, especially when it is desirable to change co-ordinates away from Cartesian type, in order to more naturally describe entropy and adiabatic perturbations. Indeed, such "co-ordinate" transformations have been done in the literature but, as shown in appendix \ref{app:twoField}, can lead to erroneous conclusions.

In this paper we shall examine a method of perturbing scalar fields proposed in \cite{AlvarezGaume:1981hn} that expresses the scalar perturbations themselves as vectors in field-space, this then allows one to write expressions that are manifestly covariant under field redefinitions. This method is then applied to the cosmological setting, and we present results for the evolution of the scalar perturbations. Such scalar perturbations are often described in terms of adiabatic and entropic perturbations, where the adiabatic perturbations desrcibe departures of the scalar along the direction of the background trajectory in field-space, and the entropy perturbations take one off the background trajectory. Therefore we propose a covariant definition of entropy and adiabatic perturbations, and give expressions for their evolution, commenting upon the current definitions in the literature. We also introduce the notion of a pseudo superpotential for multiple scalars in a cosmological setting. This is analagous to the superpotentials familiar in the supersymmetry literature, and has similar benefits to cosmology by simplifying key expressions.

In order to put the formalism to use we calculate some quantities that are of interest in cosmology: the slow-roll parameters are calculated using a Hamilton-Jacobi framework, extending previous work from single to multiple fields without the need of the slow-roll approximations; the spectral index is evaluated; and the non-Gaussianity parameter $f_{\rm NL}$ is presented in a manifestly covariant form. These are then given for the cases where the scalar potential is derived from a pseudo superpotential, allowing one to write closed-form expressions for various quantities without using the slow-roll approximation.

The paper is organized by describing the problem of scalar field perturbations in section \ref{sec:pertDef}, which are then connected to cosmological perturbations in \ref{sec:cosmoPert}. Covariant definitions of entropy and adiabatic perturbations are given in \ref{sec:entAndAdi} with pseudo supersymmetry being described in section \ref{sec:pseudosusy} and applied to the slow-roll parameters in \ref{sec:slowRollParam}. The $\delta{\cal N}$ formalism is reviewed in section \ref{sec:deltaN}, and we use this to calculate the curvature perturbation within pseudo-susy, using the field-space-covariant perturbation formulation. Various technical details are contained in the appendices.

\section{Defining the scalar-field perturbations}\label{sec:pertDef}
The essence of the problem is to find some set of quantities, $\chi^\alpha$, that will represent the perturbations $\varphi^\alpha\rightarrow\varphi^\alpha+\delta\varphi^\alpha$. A natural question to ask is, why not just use $\delta\varphi^\alpha$ as is done in various other formulations of the problem \cite{Yokoyama:2008by,astro-ph/0009131,astro-ph/9507001,Sasaki:1998ug,astro-ph/0009268,Malik:2004tf,GrootNibbelink:2000vx,GrootNibbelink:2001qt,Peterson:2010np,Peterson:2011yt}? The answer is that the perturbations $\delta\varphi^\alpha$ do not transform in a convenient way under field redefinitions $\underline\varphi\rightarrow\underline\varphi'=\underline\varphi'\left(\underline\varphi\right)$, with the transformation law being
\ba
\label{eq:deltaVariation}
\delta\varphi'^\alpha&=&\frac{\del\varphi'^\alpha}{\del\varphi^\beta}\delta\varphi^\beta+
        \frac{1}{2}\frac{\del^2\varphi'^\alpha}{\del\varphi^\beta\del\varphi^\gamma}\delta\varphi^\beta\delta\varphi^\gamma+...
\ea
So despite the use of differential geometry in some other formalisms, the base quantity is not covariant. An expression such as (\ref{eq:deltaVariation}) makes it difficult to keep track of how objects transform under field redefinitions, as well as confusing the "order" of perturbation expansions, with terms being first order in one choice of variables, but containing higher orders when written using other variables.

The key to resolving this is to use a method put forward in \cite{AlvarezGaume:1981hn}, and look for objects that describe the perturbation, but that transform covariantly as
\ba
\chi'^\alpha&=&\frac{\del\varphi'^\alpha}{\del\varphi^\beta}\chi^\beta.
\ea
Such transformation rules are of course familiar for spacetime quantities, but the same rules of covariance also apply to scalar-field manifolds. By expressing the perturbation theory in terms of tensors on the scalar-field manifold we are guaranteed to end up with physical results that are independent of how we choose to parametrize that manifold.

\begin{figure}
  \centering
  \includegraphics[width=0.5\textwidth,angle=-90]{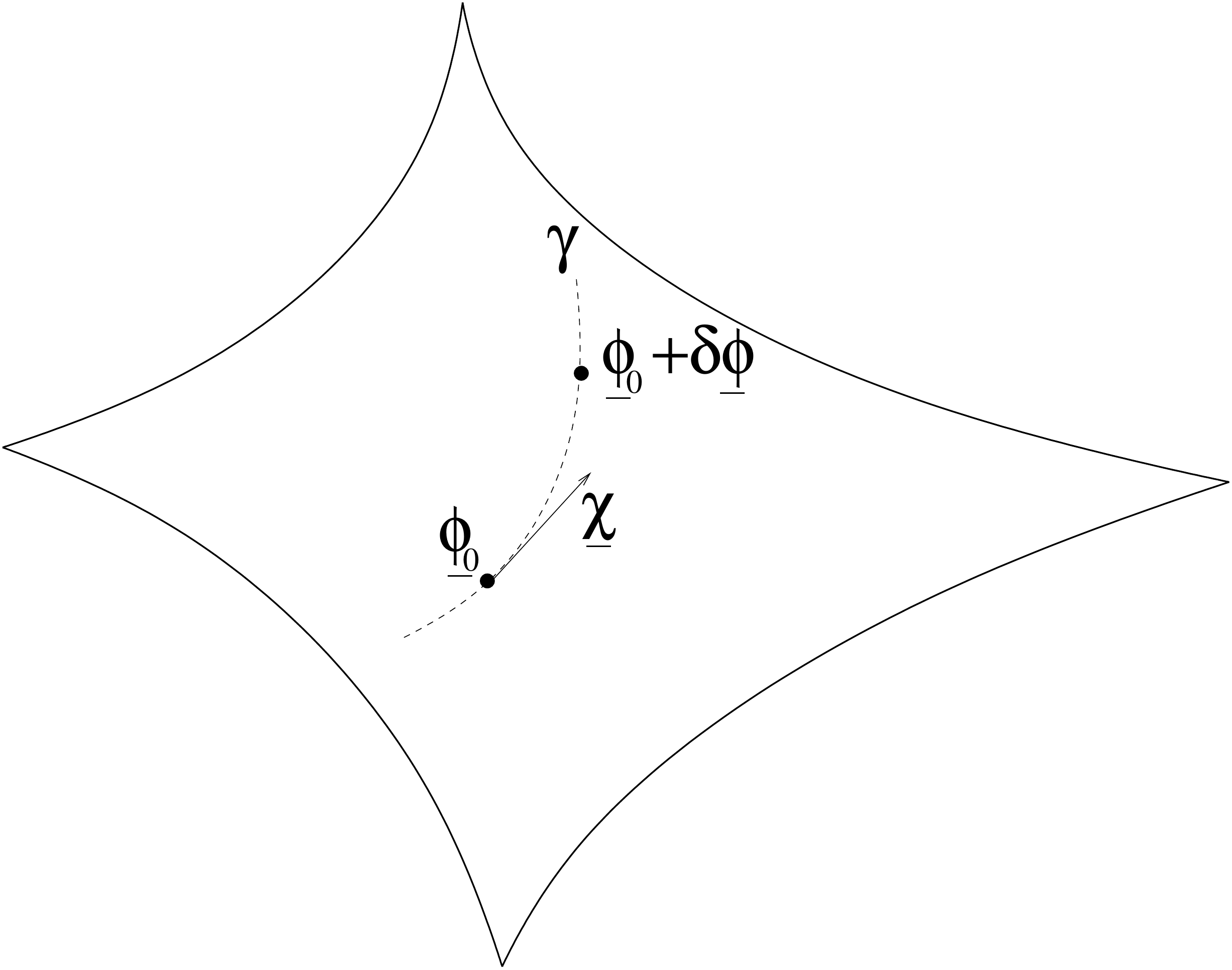}
  \caption{\label{fig:fig0}This is a plot indicating the definition of the perturbation variable $\underline\chi(=\underline\xi_0)$, as the tangent vector of the geodesic connecting the points $\varphi^\alpha_{(0)}$ and $\varphi^\alpha_{(0)}+\delta\varphi^\alpha$, evaluated at $\varphi^\alpha_{(0)}$; the dashed line represents the geodesic $\gamma$, and the norm of $\underline\chi$ is just the length of the geodesic.}
\end{figure}

The basic observation is that two nearby points on the scalar-field manifold are connected by a unique geodesic $\gamma$ (using the Christoffel connection), so we may use the tangent vector, $\underline\xi$, of such a geodesic to describe $\delta\underline\varphi$, see Fig. \ref{fig:fig0}. In this way we essentially replace $\delta\underline\varphi$ by $\underline\xi_0$, the tangent vector of $\gamma$ evaluated at the background value of the scalars, and we parametrize $\gamma$ such that $\underline\xi_0$ has norm equal to the proper length between $\underline\varphi_{(0)}$ and $\underline\varphi_{(0)}+\delta\underline\varphi$ along $\gamma$. So, our perturbation variable is now a scalar-manifold tensor, $\underline\xi_0$, which is what we were aiming for. We shall now describe how this works in practise. 


Suppose that the scalar-field manifold has metric $G_{\alpha\beta}(\underline\varphi)$\footnote{We use Greek indices $\alpha$, $\beta$,... to represent the scalar field components, and $\mu$, $\nu$,.. to represent spacetime indices.}, which we take to be Euclidean to avoid ghosts, and Christoffel symbols $\Gamma^\alpha_{\;\beta\gamma}(\underline\varphi)$, and the scalars have potential $V(\underline\varphi)$, i.e. the Lagrangian density is given by
\ba
\label{eq:scalarLagrangian}
{\cal L}&=&-\frac{1}{2}G_{\alpha\beta}(\varphi)\del_\mu\varphi^\alpha\del^\mu\varphi^\beta-V(\underline \varphi),
\ea
then the geodesic connecting $\varphi_{(0)}^\alpha$ to $\varphi_{(0)}^\alpha+\delta\varphi^\alpha$ satisfies
\ba
\label{eq:geod}
\frac{d^2\varphi^\alpha}{d\lambda^2}+\Gamma^\alpha_{\;\beta\gamma}\frac{d\varphi^\beta}{d\lambda}\frac{d\varphi^\gamma}{d\lambda}&=&0,
\ea
for some affine parameter $\lambda$ describing the journey along the path; for close enough points, as measued along the geodesic, such geodesics are unique. The tangent vector to the geodesic has components
\ba
\xi^\alpha&=&\frac{d\varphi^\alpha}{d\lambda}.
\ea
If we were to use the proper distance in field-space to parametrize the curve, 
\ba
\label{eq:dsigma}
d\sigma=\sqrt{G_{\alpha\beta}(\varphi)d\varphi^\alpha d\varphi^\beta},
\ea
 then the tangent vector, which has components $T^\alpha=\frac{d\varphi^\alpha}{d\sigma}$, would be of unit norm, $(T,T)=G_{\alpha\beta}T^\alpha T^\beta=1$. However, it is actually more convenient to choose the affine parameter $\lambda$ such that 
\ba
\varphi^\alpha(\lambda=0)&=&\varphi^\alpha_{(0)},\\
\varphi^\alpha(\lambda=1)&=&\varphi^\alpha_{(0)}+\delta\varphi^\alpha.
\ea
The reason for this, as we shall see, is that then the norm of the perturbation variable will just be given by the proper distance in field-space between $\varphi^\alpha_{(0)}$ and $\varphi^\alpha_{(0)}+\delta\varphi^\alpha$.

We now define the Riemann co-ordinates of some point near $\varphi^\alpha_{(0)}$ to be $\tilde\varphi^\alpha$, given by
\ba
\tilde\varphi^\alpha&=&\sigma T^\alpha_{(0)}=\lambda\xi^\alpha_{(0)},
\ea
where $\sigma$ is the proper distance to the point from $\varphi^\alpha_{(0)}$ along the connecting geodesic. This implies\footnote{We use $(\;,)$ to denote the inner product, using the metric components $G_{\alpha\beta}$, $(A,B)=G_{\alpha\beta}A^\alpha B^\beta$.}
\ba
\sigma^2(T_{(0)},T_{(0)})&=&\lambda^2(\xi_{(0)},\xi_{(0}),
\ea
and so if we recall that the $T^\alpha$ describe a unit norm tangent vector, then at $\lambda=1$ we have
\ba
(\xi_{(0)},\xi_{(0)})=\sigma^2.
\ea
For this reason we define our perturbation variable, $\chi^\alpha$, to be
\ba
\label{eq:chiDef}
\chi^\alpha=\xi^\alpha_{(0)}.
\ea

It is these variables that play the role of the $\delta\varphi^\alpha$; they transform covariantly as they are the components of a tangent vector, and they have a norm equal to the proper distance between the background field and the perturbed field. Simply put, if we want to go from $\varphi^\alpha_{(0)}$ to $\varphi^\alpha_{(0)}+\delta\varphi^\alpha$, just go along the geodesic associated to the tangent vector $\xi^\alpha_{(0)}$, Fig. \ref{fig:fig0}.

Combining everything together, the perturbations of $\underline\varphi$, $G_{\alpha\beta}(\underline\varphi)$ and $V(\underline\varphi)$, from appendix \ref{app:scalarFieldPert} we find the action for the scalar field perturbations $\chi^\alpha$ to be\footnote{This corrects some typos in \cite{AlvarezGaume:1981hn}.}
\ba
\label{eq:perturbedAction}
&~&S[\varphi+\delta\varphi]\\\nonumber
&=&\int d^4x\sqrt{-g}\left[\left(-\frac{1}{2}G_{\alpha\beta}\del_\mu\varphi^\alpha\del^\mu\varphi^\beta-V\right)
  +\left(-G_{\alpha\beta}\del_\mu\varphi^\alpha D^\mu\chi^\beta-\del_\alpha V\chi^\alpha\right)\right.\\\nonumber
     &~&-\frac{1}{2}\left( G_{\alpha\beta}D_\mu\chi^\alpha D^\mu\chi^\beta
                          +R_{\beta\alpha_1\alpha_2\gamma}\chi^{\alpha_1}\chi^{\alpha_2}\del_\mu\varphi^\beta\del^\mu\varphi^\gamma
                          +D_\alpha\del_\beta V\chi^\alpha\chi^\beta\right)\\\nonumber
   &~&-\frac{1}{3!}\left( D_{\alpha_1}R_{\beta\alpha_2\alpha_3\gamma}\chi^{\alpha_1}\chi^{\alpha_2}\chi^{\alpha_3}\del_\mu\varphi^\beta\del^\mu\varphi^\gamma
             +4R_{\beta\alpha_1\alpha_2\alpha_3}\chi^{\alpha_1}\chi^{\alpha_2}D_\mu\chi^{\alpha_3}\del^\mu\varphi^\beta
             +D_{\alpha_1}D_{\alpha_2}\del_{\alpha_3}V\chi^{\alpha_1}\chi^{\alpha_2}\chi^{\alpha_3}\right)\\\nonumber
   &~&-\frac{1}{4!}\left( 
          \left[ D_{\alpha_1}D_{\alpha_2}R_{\beta\alpha_3\alpha_4\gamma}
                   +4R_{\;\beta\alpha_1\alpha_2\delta}R^\delta_{\;\alpha_3\alpha_4\gamma}\right]\chi^{\alpha_1}\chi^{\alpha_2}\chi^{\alpha_3}\chi^{\alpha_4}\del_\mu\varphi^\beta\del^\mu\varphi^\gamma\right.\\\nonumber
  &~&\;\;\;\;\;\;\;\;+6D_{\alpha_1}R_{\beta\alpha_2\alpha_3\alpha_4}\chi^{\alpha_1}\chi^{\alpha_2}\chi^{\alpha_3}D_\mu\chi^{\alpha_4}\del^\mu\varphi^\beta
             +4R_{\alpha_1\alpha_2\alpha_3\alpha_4}D_\mu\chi^{\alpha_1}\chi^{\alpha_2}\chi^{\alpha_3}D^\mu\chi^{\alpha_4}\\\nonumber
          &~&\;\;\;\;\;\;\;\;\left.\left.+D_{\alpha_1}D_{\alpha_2}D_{\alpha_3}\del_{\alpha_4}V\chi^{\alpha_1}\chi^{\alpha_2}\chi^{\alpha_3}\chi^{\alpha_4}\right)+...\right],\\\nonumber
\ea
where $D_\alpha$ is the covariant derivative on the scalar manifold, using the Christoffel connection, and $R^\alpha_{\;\;\beta\gamma\delta}$ is its Riemann curvature. In this expression we are now conforming to the notation of \cite{AlvarezGaume:1981hn} by dropping the zero subscript on $\underline\varphi$, which is now used to denote the background field. This is now a {\it fully} covariant expression, both in spacetime, and field-space, with the $\underline\chi$ being spacetime scalars, and field-space vectors.

\section{Cosmological scalar perturbations}\label{sec:cosmoPert}
In this section we shall revisit some earlier calculations of multi-field scalar perturbations \cite{astro-ph/0603799,astro-ph/0009268,astro-ph/0211602,astro-ph/0009131,arXiv:1008.3198,astro-ph/9507001,Sasaki:1998ug,astro-ph/0506056} with an emphasis on writing expressions that are manifestly covariant in field-space.

The energy momentum tensor and equations of motion following from (\ref{eq:scalarLagrangian}) allows us to determine the background quantities of energy density, pressure and total adiabatic sound speed
\ba
\label{eq:rhoBar}
\bar\rho&=&\frac{1}{2}(\dot\varphi,\dot\varphi)+V,\\
\label{eq:Pbar}
\bar P&=&\frac{1}{2}(\dot\varphi,\dot\varphi)-V,\\
\label{eq:soundSpeed}
c_s^2&=&\dot{\bar P}/\dot{\bar \rho}=1+\frac{2(\del V,\dot\varphi)}{3H(\dot\varphi,\dot\varphi)},
\ea
where a bar over a quantity indicates it is a background quantity, and $H$ is the Hubble parameter $H=\frac{1}{a}\frac{da}{dt}$.

We may also, using (\ref{eq:perturbedFluidVelocity}, \ref{eq:fluidEMtensor}), evaluate the perturbation of the energy density, pressure and fluid three-velocity
\ba
\delta\rho&=&(\dot\varphi, D_t\chi)-\phi (\dot\varphi,\dot\varphi)+(\del V,\chi),\\
\delta P&=&(\dot\varphi,D_t\chi)-\phi(\dot\varphi,\dot\varphi)-(\del V,\chi),\\
\label{eq:rhoBarPlusPbar}
(\bar\rho+\bar P)av&=&-(\dot\varphi,\chi),
\ea
Then, using (\ref{eq:rhoBarPlusPbar}, \ref{eq:kappaRel}, \ref{eq:kappaDef}) we find a relation between the metric potentials and our scalar field perturbation
\ba
\label{eq:Hphi}
H\phi+\dot\psi+\frac{K\sigma_s}{a^2}&=&4\pi G(\dot\varphi,\chi).
\ea
The action (\ref{eq:scalarLagrangian}) leads to the equation of motion for the background scalar-field 
\ba
{\cal D}_\mu \del^\mu\varphi^\alpha-G^{\alpha\beta}\del_\beta V&=&0,
\ea
where we remind the reader that $\del_\mu\varphi^\alpha$ transforms as the components of a spacetime vector, and a field-space vector, so its derivative must be covariant both in spacetime ($\nabla$) and field-space ($D$), which together we write as ${\cal D}$ (\ref{eq:spacetimeFieldCovDer}). On an FRW background this becomes
\ba
\label{eq:phiEOM}
D_t \dot\varphi^\alpha+3H\dot\varphi^\alpha+G^{\alpha\beta}\del_\beta V=0.
\ea
The action for the scalar-field perturbation (\ref{eq:perturbedAction}), at quadratic order, leads to 
\ba\nonumber
&~&D_tD_t\chi^\alpha+3HD_t\chi^\alpha-\frac{1}{a^2}\Delta\chi^\alpha-R^\alpha_{\;\;\beta\gamma\delta}\dot\varphi^\beta\dot\varphi^\gamma\chi^\delta+D^\alpha\del_\beta V\chi^\beta\\
    &~&\qquad\qquad =(\kappa+\dot\phi)\dot\varphi^\alpha+\left[2D_t\dot\varphi^\alpha+3H\dot\varphi^\alpha\right]\phi
\ea
which, using (\ref{eq:Hkappa},\ref{eq:rhoBar},\ref{eq:Hdot},\ref{eq:Hphi}) and taking $K=0$, may be written in the form analogous to that in \cite{astro-ph/9507001} 
\ba
\nonumber
\label{eq:evolvePert}
&~&D_tD_t\chi^\alpha+3HD_t\chi^\alpha_{(\psi)}-\frac{1}{a^2}\Delta\chi^\alpha
    -R^\alpha_{\;\;\beta\gamma\delta}\dot\varphi^\beta\dot\varphi^\gamma\chi^\delta+D^\alpha\del_\beta V\chi^\beta\\
    &~&\qquad\qquad =\frac{8\pi G}{a^3}D_t\left(\frac{a^3}{H}G_{\beta\gamma}\dot\varphi^\alpha\dot\varphi^\beta\right)\chi^\gamma.
\ea
In deriving this it is simplest to use the gauge $\psi=0$ (\ref{eq:perturbedLineElement}), but it is a simple matter to express it in gauge invariant variables, simply replace $\chi^\alpha$ with $\chi^\alpha_{(\psi)}$ as defined in App. \ref{app:gaugeInvVar}. 

\subsection{Consistent entropy and adiabatic perturbations}\label{sec:entAndAdi}
In order to picture the perturbations it is sometimes convenient to split them into those parallel to the background evolution of $\underline\varphi$ (adiabatic perturbations), and those normal to it (entropy perturbations); as such, we need to define the split. Our first observation is that a common definition for the entropy perturbations \cite{astro-ph/0009268}\cite{Malik:2004tf} $\sim\frac{\delta\varphi^\alpha}{\dot\varphi^\beta}-\frac{\delta\varphi^\beta}{\dot\varphi^\alpha}$ will not do, they are very non-covariant objects. Nor can one split up the components of $\underline\varphi$ and define density perturbations for each component of the form $\dot\varphi^\alpha\delta\dot\varphi^\alpha-(\dot\varphi^{\alpha})^2\phi$, $\del V_\alpha\delta\varphi^\alpha$ (no sum over $\alpha$)\cite{Malik:2004tf}, as this is not covariant either. In the two-field case Gordon {\it et al} \cite{astro-ph/0009131} avoided this issue by explicitly working with the components, however, that introduces a problem of its own (see App. \ref{app:twoField}). Here we propose the following covariant definitions for the adiabatic perturbation $\delta\sigma$ and the entropy perturbations $\delta S^{\alpha\beta}$,
\ba
\label{eq:deltaSigmaDef}
\delta\sigma&=&\frac{(\dot\varphi,\chi)}{\dot\sigma},\\
\label{eq:deltaSDef}
\delta S^{\alpha\beta}&=&2\frac{\dot\varphi^{[\alpha}\chi^{\beta]}}{\dot\sigma},
\ea
where the antisymmetrization is defined by $[xy]=\frac{1}{2}(xy-yx)$. As $\underline{\dot\varphi}$ is the tangent vector of the background trajectory then $\delta\sigma$ is just the component of the perturbation along the direction of the background evolution, with $\delta S$ representing those orthogonal to it. Note that we may invert these relations to recover the $\chi$ perturbation from the entropy and adiabatic perturbations using 
\ba
\label{eq:invertEntAd}
\chi^\alpha=\frac{1}{\dot\sigma}\dot\varphi^\alpha\delta\sigma-\frac{1}{\dot\sigma}\delta S^{\alpha\beta}\dot\varphi_\beta.
\ea
If we were to use $\delta\varphi^\alpha$ instead of $\chi^\alpha$ then these would agree with the two-field case examined in \cite{astro-ph/0009131}, however, it is important to note that $(\varphi^1,\varphi^2)\rightarrow(\sigma,S)$ does not in general constitute a field redefinition, and assuming it does leads to incorrect equations for the evolution of perturbations \cite{astro-ph/0009131} (App.\ref{app:twoField}). After some work one finds that these new variables evolve according to
\ba
&~&\ddot\delta\sigma+3H\dot\delta\sigma-\frac{1}{a^2}\Delta\delta\sigma
+\frac{1}{\dot\sigma^2}\left[(V_{,\sigma})^2-(\del V,\del V)\right]\delta\sigma
+\frac{1}{\dot\sigma^2}D_\alpha\del_\beta V\dot\varphi^\alpha\dot\varphi^\beta\delta\sigma\\\nonumber
&=& \frac{2}{\dot\sigma^3}V_{,\sigma}\delta S^{\alpha\beta}\del_\alpha V\dot\varphi_\beta
   +\frac{2}{\dot\sigma^2}\delta S^{\alpha\beta}D_\gamma\del_\alpha V\dot\varphi^\gamma\dot\varphi_\beta
   +\frac{2}{\dot\sigma^2}D_t\delta S^{\alpha\beta}\del_\alpha V\dot\varphi_\beta\\\nonumber
 &~&+\frac{8\pi G}{H}\left[\frac{2}{\dot\sigma^2}(\dot\sigma^2+V)\delta S^{\alpha\beta}\del_\alpha V\dot\varphi_\beta
                          -2\dot\sigma V_{,\sigma}\delta\sigma
                          -\frac{8\pi G}{H}\dot\sigma^2 V\delta\sigma\right],
\ea
and\footnote{
Here we define $V_{,\sigma}=\dot\varphi^\alpha\del_\alpha V$ and 
$V_{,\sigma\sigma}=\dot V_{,\sigma}/\dot\sigma=D_\alpha\del_\beta V\varphi^\alpha\dot\varphi^\beta/\dot\sigma^2+[V_{,\sigma}^2-(\del V,\del V)]/\dot\sigma^2$.
}
\ba
&~&D_tD_t\delta S^{\alpha\beta}+3HD_t\delta S^{\alpha\beta}-\frac{1}{a^2}\Delta\delta S^{\alpha\beta}\\\nonumber
&=& \frac{2}{\dot\sigma}V_{,\sigma}\left(D_t\delta S^{\alpha\beta}+3H\delta S^{\alpha\beta}\right)
   +\frac{4}{\dot\sigma^2}\del^{[\alpha}V\left(D_t\delta S^{\beta]\gamma}+3H\delta S^{\beta]\gamma}\right)\dot\varphi_\gamma\\\nonumber
&~&+\frac{4}{\dot\sigma^3}V_{,\sigma}\del^{[\alpha}V\delta S^{\beta]\gamma}\dot\varphi_\gamma
   -\frac{4}{\dot\sigma^2}\del^{[\alpha}V\delta S^{\beta]\gamma}\del_\gamma V\\\nonumber
&~&+\frac{2}{\dot\sigma^2}D_\delta\del^{[\alpha}V\delta S^{\beta]\gamma}\dot\varphi^\delta\dot\varphi_\gamma
   +\frac{2}{\dot\sigma^2}\dot\varphi^{[\alpha}D^{\beta]}\del_\gamma V\delta S^{\gamma\delta}\dot\varphi_\delta
   +V_{,\sigma\sigma}\delta S^{\alpha\beta}\\\nonumber
&~&-\frac{2}{\dot\sigma^2}\dot\varphi^{[\alpha}R^{\beta]}_{\;\;\gamma\delta\epsilon}\dot\varphi^\gamma\dot\varphi^\delta\delta S^{\epsilon\eta}\dot\varphi_\eta\\\nonumber
&~&-\frac{4}{\dot\sigma^3}(3H\dot\sigma+V_{,\sigma})\del^{[\alpha}V\dot\varphi^{\beta]}\delta\sigma
   +\frac{16\pi G}{H}\del^{[\alpha}V\dot\varphi^{\beta]}\delta\sigma-\frac{4}{\dot\sigma^2}\del^{[\alpha}V\dot\varphi^{\beta]}\delta\dot\sigma
\ea
In practise it may be easier to evolve the system of perturbations with (\ref{eq:evolvePert}) and simply use (\ref{eq:deltaSigmaDef}), (\ref{eq:deltaSDef}) to evaluate the entropy and adiabatic components, with (\ref{eq:invertEntAd}) allowing us to set initial conditions for perturbations in terms of adiabatic and entropy modes. However there may be situations where one must evolve the entropy and adiabatic perturbations due to the finite precision of numerical methods\cite{wandsPrivComm}.
\section{Pseudo-susy and multi-field slow-roll parameters}\label{sec:pseudosusy}
In order to use some of this formalism we are going to extend an idea that was used to study dark energy with a single scalar field \cite{Bazeia:2007vx}, and is also useful in understanding the similarity between the equations describing a gravitating domain wall, and the cosmology of evolving scalar fields \cite{Bazeia:2005tj,Skenderis:2006fb,McFadden:2009fg}. It has also been extended to multiple scalars \cite{Chemissany:2007fg} to examine the role of geodesics in field-space and cosmological evolution\cite{Karthauser:2006ix}. Pseudo-supersymmetry will allow us to make exact statements about the slow-roll parameters, as well as exact relations in the $\delta {\cal N}$ formalism for curvature perturbations that lead to the the spectral index and the non-Gaussianity parameter $f_{\rm NL}$. The basic observation is that the background system of equations (\ref{eq:rhoBar},\ref{eq:Pbar},\ref{eq:phiEOM},\ref{eq:friedmann},\ref{eq:Hdot})
\ba
\label{eq:Fried}
3H^2&=&8\pi G\left(\frac{1}{2}(\dot\varphi,\dot\varphi)+V\right),\\
\label{eq:Hdot2}
\dot H&=&-4\pi G(\dot\varphi,\dot\varphi),\\
\label{eq:phi_eom}
D_t \dot\varphi^\alpha&=&-3H\dot\varphi^\alpha-\del^\alpha V,
\ea
where we have set the spatial curvature $K$ to zero, may be solved by the following system
\ba
\label{eq:gradFlow}
\dot\varphi^\alpha&=&\pm\del^\alpha W,\\
\label{eq:HW}
H&=&\mp4\pi GW,
\ea
if we impose a special form on the potential, namely\footnote{It is worth pointing out that the calculation of \cite{Byrnes:2009qy} can be recast using this framework.}
\ba
\label{eq:cosmosusy}
V(\underline\varphi)=6\pi GW^2-\frac{1}{2}(\del W,\del W).
\ea
This is rather like the BPS-system one finds for certain domain-walls \cite{Chamblin:1999cj} and, in line with that setup, we shall refer to $W(\underline\varphi)$ as the pseudo superpotential\footnote{In the supergravity literature $W$ is usually reserved for the superpotential and $V$ the potential, whereas $W$ is often used in the cosmology literature to denote the potential; we shall use the supergravity-style convention.}. This requirement has a remarkable simplification for the slow-roll parameters, defined in terms of the Hamilton-Jacobi system, as we shall now demonstrate.

The Hamilton-Jacobi formalism has been examined for single scalars \cite{Liddle:1994dx}\cite{Kinney:1997ne}, see also \cite{Lyth:2009zz}, and comes about by thinking of the Hubble parameter as a function of the single scalar field, $\Phi$. 
The first two slow-roll parameters are then defined by
\ba
\epsilon_H&=&\frac{1}{4\pi G}\left(\frac{H'(\Phi)}{H(\Phi)}\right)^2,\\
\eta_H&=&\frac{1}{4\pi G}\left(\frac{H''(\Phi)}{H(\Phi)}\right),
\ea
where $'$ denotes differentiation with respect to $\Phi$; higher order parameters are given in \cite{Liddle:1994dx}.

The case of multi-scalar fields is more complicated and there are a number of different ways to generalize the slow-roll parameters \cite{GrootNibbelink:2000vx,GrootNibbelink:2001qt,Peterson:2010np,Peterson:2011yt,Burgess:2004kv,BlancoPillado:2006he,Chiba:2008rp}, here we propose slow-roll parameters that are rooted in the Hamilton-Jacobi formalism. The essential point to note is that the background evolution picks out a particular path in field space, and it is really only that path that plays any role in background evolution. As such, what we are interested in is how $H$ varies as $\sigma$ (the proper distance along the path in field-space) increases. A few important relations coming from (\ref{eq:Fried}), (\ref{eq:Hdot2}), (\ref{eq:phi_eom}) are
\ba
\label{eq:Hsigma}
H_{,\sigma}&=&-4\pi G\dot\sigma,\\
(H_{,\sigma})^2-12\pi GH^2&=&-2(4\pi G)^2V,\\
\ddot\sigma+3H\dot\sigma+V_{,\sigma}&=&0,
\ea
with the $\sigma$ derivative being defined by
\ba
H_{,\sigma}=\frac{\del_\alpha H\dot\varphi^\alpha}{\dot\sigma}.
\ea

In the multi-field case we therefore define
\ba
\epsilon_H&=&\frac{1}{4\pi G}\left(\frac{H_{,\sigma}}{H}\right)^2,\\
\eta_H&=&\frac{1}{4\pi G}\left(\frac{H_{,\sigma\sigma}}{H}\right),
\ea
and we note that, just as in the single-field case, (\ref{eq:Fried}), (\ref{eq:Hdot2}) and (\ref{eq:Hsigma}) lead to\footnote{Recall that $(\dot\varphi,\dot\varphi)=\dot\sigma^2$ (\ref{eq:dsigma}).}
\ba
\frac{\ddot a}{a}&=&H^2(1-\epsilon_H),
\ea
showing that inflationary solutions have $\epsilon_H<1$.
\subsection{Slow-roll parameters in the slow-roll limit}
\label{sec:slowRollParam}
In the slow-roll approximation we neglect $D_t\dot\varphi^\alpha$ compared to $H\dot\varphi^\alpha$ in the scalar equation of motion (\ref{eq:phi_eom}), and we drop $(\dot\varphi,\dot\varphi)$ when compared to $V$ in the Friedmann equation (\ref{eq:Fried}), giving that
\ba
\epsilon_H\simeq\frac{1}{16\pi G}\frac{(\del V,\del V)}{V^2},
\ea
which is a standard result, and
\ba
\eta_H&\simeq&\epsilon_H-\frac{1}{8\pi G}\frac{D_\alpha\del_\beta V}{V}\frac{\del^\alpha V\del^\beta V}{(\del V,\del V)},
\ea
which reduces to the standard expression for the single-field case, and recovers the expression in \cite{Lyth:2009zz} when we take the field-metric to be flat and written in Cartesian co-ordinates. We may also reduce the expression for the number of e-folds of inflation to the standard approximate integral expression
\ba
N&=&\int dt H=\int dt\frac{H^2}{H}\simeq\frac{8\pi G}{3}\int dt\frac{V}{H}\simeq8\pi G\int d\varphi^\alpha\frac{V}{3H\dot\varphi^\alpha}\\\nonumber
   &\simeq&-8\pi G\int d\varphi^\alpha\frac{V}{\del^\alpha V},
\ea
where there is no sum over the $\alpha$\footnote{The apparent non-covariant nature of this expression is removed by requiring it to be evaluated on-slow-roll-shell, $3H\dot\varphi^\alpha+\del^\alpha V=0$.}.

\subsection{Slow-roll parameters without the slow-roll}
\label{sec:slowRollParamNo}
The remarkable thing about pseudo-susy (\ref{eq:cosmosusy}) is that we now get expressions for the slow-roll parameters in terms of the pseudo superpotential without the need for making the slow-roll approximation,
\ba
\epsilon_H&=&\frac{1}{4\pi G}\frac{(\del W,\del W)}{W^2},\\
\eta_H&=&\frac{1}{4\pi G}\frac{D_\alpha\del_\beta W}{W}\frac{\del^\alpha W\del^\beta W}{(\del W,\del W)},
\ea
and the expression for the number of e-folds is now an integral expression that does not require the slow-roll approximation
\ba
\nonumber
\label{eq:e-foldNumber}
N&=&\int dt\;H=\mp 4\pi G \int\; dt\;W=\mp4\pi G \int \; dt\;(W-Q)\mp4\pi G \int\; dt\;Q\\
 &=&-4\pi G\int d\varphi^\alpha\frac{W-Q}{\del^\alpha W}-4\pi G\int d\varphi^\beta\frac{Q}{\del^\beta W},
\ea
where $Q$ is some function that has been added and subtracted in order to make the integrals solvable - in some cases. 

With an eye on systems that we are able to solve we now give some examples of pseudo superpotentials
\begin{itemize}
\item
CLASS I: generalized  sum-separable 
\ba
W&=&\left[w_{(1)}(\varphi^1)+w_{(2)}(\varphi^2)+w_{(3)}(\varphi^3)+...\right]^m,\\
\frac{N}{4\pi G}&=&-\sum_\beta\int\;d\varphi^\beta\frac{w_{(\beta)}}{mw_{(\beta)}'}.
\ea
\item
CLASS II: product-separable 
\ba
W&=&w_1(\varphi^1)w_2(\varphi^2)w_3(\varphi^3)...,\\
\frac{N^{(\beta)}}{4\pi G}&=&-\int\;d\varphi^{\beta}\frac{w_{(\beta)}}{w_{(\beta)}'}.
\ea
\end{itemize}
The form of these stems from the analogous systems in the slow-roll limit using the potential rather than our pseudo superpotentials  \cite{astro-ph/0603799,arXiv:1008.3198,Battefeld:2006sz,Choi:2007su,GarciaBellido:1995qq}. The slow-roll approximation using $V$ yields similar expressions for $N$ \cite{arXiv:1008.3198}, but in the following we shall be using pseudo-susy allowing us to derive expressions without the need for such an approximation.

\section{$\delta {\cal N}$, $n_\zeta$ and f$_{NL}$}\label{sec:deltaN}
In this section we shall combine our field-space covariant approach to scalar perturbations with pseudo-susy to give expressions for the spectral index and the non-Gaussianity parameter, which does not rely on making the slow-roll approximation. First we give a recap of the $\delta{\cal N}$ formalism \cite{Starobinsky:1986fxa,astro-ph/9507001,Sasaki:1998ug,Lyth:2004gb,astro-ph/0506262}.

The important idea is that the curvature perturbation on uniform energy density hypersurfaces, $\zeta$, is, on large scales, just given by the perturbation in the number of e-folds,
\ba
\zeta(t_c,\underline x)&\simeq&\delta {\cal N}(t_c,t_\star,\underline x)\equiv{\cal N}(t_c,t_\star,\underline x)-N(t_c,t_\star),
\ea
where
\ba
N(t_c,t_\star)=\int_\star^c H\;dt,
\ea
is the unperturbed version of ${\cal N}$, which itself is the integral of the volume expansion rate \cite{astro-ph/9507001}. The initial hypersurface, at $t=t_\star$, is taken to be spatially flat, and the final hypersurface, at $t=t_c$, is a considered to be a uniform density hypersurface. Now we view the number of e-foldings, ${\cal N}(t_c,t_\star,\underline x)$ as depending upon the value of the scalar fields at the initial hypersurface, $\varphi(t_\star,\underline x)$, and also depending upon $t_c$. The variations $\delta {\cal N}$ may now be given in terms of the variations of $\underline\varphi_\star$. However, the standard expression \cite{astro-ph/0603799,Lyth:2005fi}
\ba
\label{eq:deltaN}
\delta{\cal N}(t_c,t_\star,\underline x)&=&\del_\alpha N\delta\varphi^\alpha_\star+\del_\alpha\del_\beta N\delta\varphi^\alpha_\star\delta\varphi^\beta_\star+...,
\ea
although correct, does not fit our ethos of writing expressions that are covariant under field redefinition - due to the non-covariant transformation properties of partial derivatives and $\delta\varphi^\alpha$. In fact, the partial derivatives appearing in expressions of the form (\ref{eq:2ndDerivN}) have this Taylor expansion as their origin, and it is the non-trivial transformation properties of $\delta\varphi^\alpha$ and $\del_\alpha$ that lead to inconsistent expressions such as (\ref{eq:2ndDerivN}). Following the procedure that led to (\ref{eq:pertPot}) we find that in our perturbation variables
\ba
\label{eq:deltaNexp}
\delta{\cal N}(t_c,t_\star,\underline x)&=&\del_\alpha N\chi^\alpha_\star+D_\alpha\del_\beta N\chi^\alpha_\star\chi^\beta_\star+...
\ea
which is manifestly covariant, now all we need to do is calculate it.
\subsection{$n_\zeta$ and $f_{\rm NL}$}
Following \cite{astro-ph/0603799} we have that the curvature perturbation power spectrum ${\cal P}_\zeta$ is defined by
\ba
\langle\zeta(\underline k_1)\zeta(\underline k_2)\rangle&=&(2\pi)^3\delta_3(\underline k_1+\underline k_2)\frac{2\pi^2}{k_1^3}{\cal P}_\zeta(k_1),
\ea
and we want to relate this to the two-point correlator of the scalar field perturbations. Again, we see that the standard expression $\langle\delta\varphi^\alpha(\underline k_1)\delta\varphi^\beta(\underline k_2)\rangle\sim G^{\alpha\beta}$ \cite{astro-ph/0603799,Yokoyama:2008by,astro-ph/0009131,astro-ph/0506056} cannot be true because the left and right hand sides transform differently under field redefinition. Instead we have that 
\ba
\label{eq:twoPointScalar}
\langle\chi^\alpha(\underline k_1)\chi^\beta(\underline k_2)\rangle&=&
    (2\pi)^3G^{\alpha\beta}\delta_3(\underline k_1+\underline k_2)\frac{2\pi^2}{k_1^3}{\cal P}_\star(k_1),\\
{\cal P}_\star(k_1)&=&\frac{H^2_\star}{4\pi^2},
\ea
where $H_\star$ is evaluated at $k=aH$ \footnote{see \cite{AlvarezGaume:1981hn} for the flat spacetime version of (\ref{eq:twoPointScalar}).}. Now we combine these with (\ref{eq:deltaN})(\ref{eq:deltaNexp}) to find \cite{astro-ph/0603799}
\ba
{\cal P}_\zeta&=&(\del N,\del N){\cal P}_\star.
\ea
From these ingredients we may now derive the spectral index, $n_\zeta$, defined by
\ba
n_\zeta-1&=&\frac{d\ln {\cal P}_\zeta}{d\ln k}.
\ea
For inflation, this is simplified by noting that at Hubble exit, $k=aH$, and $H$ is approximately constant so \cite{Lyth:1998xn}
\ba
n_\zeta-1&\simeq&\frac{1}{H}\frac{d\ln{\cal P}_\zeta}{dt},
\ea
which we evaluate to 
\ba
\label{eq:nzeta}
n_\zeta-1&\simeq&-2\epsilon_H+\frac{2}{H}\frac{\dot\varphi^\alpha D_\alpha\del_\beta N\del^\beta N}{(\del N,\del N)},
\ea
and this is just the covariant version of what appears in \cite{astro-ph/0603799}. To get the covariant version of $f_{nl}$ we may follow the proceedure in \cite{astro-ph/0603799} using our variables to find
\ba
\label{eq:fnl}
-\frac{6}{5}f_{nl}^{(4)}&=&\frac{\del^\alpha N D_\alpha\del_\beta N\del^\beta N}{(\del N,\del N)^2}.
\ea
\subsection{Derivatives of $N$}
In the preceding sections we saw that the various formulae required derivatives of $N$ with respect to the scalar fields. We shall follow the method of Vernizzi and Wands \cite{astro-ph/0603799}, but instead of applying the slow-roll approximations using $V$, we shall apply pseudo-susy using $W$ to avoid those approximations. 

\begin{figure}
  \centering
  \includegraphics[width=0.5\textwidth,angle=-90]{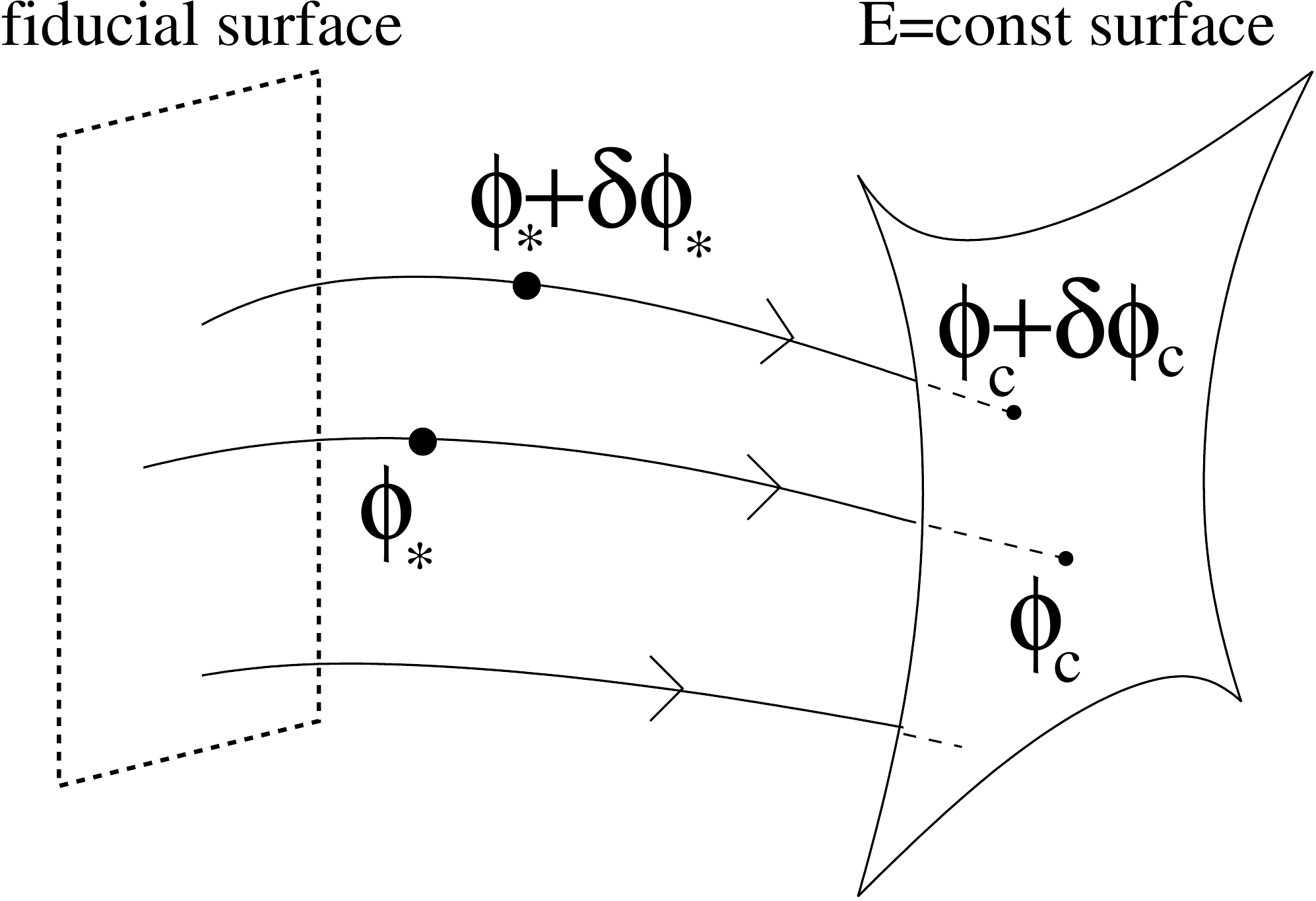}
  \caption{\label{fig:fig1}Flow lines following (\ref{eq:gradFlow}) are specified purely in terms of the gradient of the superpotential and so do not cross, except at critical point of $W$. As such they may be specified by their intersection with some fiducial co-dimension one surface. These flow lines then proceed to the constant energy density surface at $t_c$.}
\end{figure}

We may picture the evolution of the background scalar fields as some curve in field-space. Moreover, if we work with the gradient-flow solutions of pseudo-susy (\ref{eq:gradFlow}) then these flow lines do not intersect. In that case we have that the curves are parametrized by $n_\varphi-1$ constants, which may be thought of as the location of the intersection of the curves with some fiducial co-dimension one surface in field-space, Fig. \ref{fig:fig1}. In practise there is a more convenient way to parametrize the curves, which we see by noting that  the gradient-flow evolution of (\ref{eq:gradFlow}) leads to 
\ba
\label{eq:exactDiff}
\del^\alpha W\;d\varphi^\beta-\del^\beta W\;d\varphi^\alpha&=&0
\ea
which may, in principle, be integrated along the curves, so allowing one to assign a set of constants to the curves. At this point we restrict ourselves to the two classes of superpotential given in section \ref{sec:slowRollParamNo}, as this will enable us to integrate (\ref{eq:exactDiff})

Now we pick some pair $(\alpha,\beta)$ of components and suppose that we can write
\ba
\del^\alpha W&=&{\cal F}_{(\alpha,\beta)}(\underline\varphi)f_{(\beta)}(\varphi^\beta),\\
\del^\beta W&=&{\cal F}_{(\alpha,\beta)}(\underline\varphi)f_{(\alpha)}(\varphi^\alpha),
\ea
such that
\ba
{\cal F}_{(\alpha,\beta)}\left(f_{(\beta)}(\varphi^\beta)\;d\varphi^\beta-f_{(\alpha)}(\varphi^\alpha)\;d\varphi^\alpha\right)&=&0,
\ea
i.e. we extract a common function from each of the terms in (\ref{eq:exactDiff}) and ensure that the $d\varphi^\alpha$ differential is multiplied by a function of $\varphi^\alpha$, whilst the $d\varphi^\beta$ differential is multiplied by a function of $\varphi^\beta$. This includes (for flat moduli metric in Cartesian field-space co-ordinates) our CLASS I, II pseudo superpotentials given above.
Given this restriction we may now introduce a set of constants that specify a given trajectory
\ba
\label{eq:thePathConstants}
\hat C^{\alpha,\beta}&=&\int f_{(\alpha)}(\varphi^\alpha)d\varphi^\alpha-\int f_{(\beta)}(\varphi^\beta)d\varphi^\beta.
\ea
We also note that this does indeed just give $n_\varphi-1$ independent constants as, for example, $\hat C^{1,5}=\hat C^{1,2}+\hat C^{2,3}+\hat C^{3,4}+\hat C^{4,5}$. Indeed, we may take $\hat C^{1,2}$, $\hat C^{2,3}$, $\hat C^{3,4}$,...$\hat C^{n_\varphi-1,n_\varphi}$,  as our independent constants and it is then convenient, at times, to denote the independent constants by 
\ba
C^\alpha&=&\hat C^{\alpha,\alpha+1},\qquad \alpha=1...n_\varphi-1.
\ea

Now let us recall what we need to do. Equations coming from the $\delta{\cal N}$ formalism, such as (\ref{eq:nzeta},\ref{eq:fnl}), require us to differentiate $N$ with respect to $\varphi^\alpha_\star$, taking into account that as we vary $\varphi^\alpha_\star$, $\varphi^\alpha_c$ will also change as we will have moved onto another trajectory. Schematically then, we have from (\ref{eq:e-foldNumber}) that 
\ba
N&\sim&\int_\star^c\;d\varphi\frac{W}{W'},\\
dN&\sim& \left.\frac{W}{W'}\right|_\star d\varphi_\star-\left.\frac{W}{W'}\right|_c \frac{\del\varphi_\star}{\del\varphi_c}d\varphi_\star,
\ea
with all the hard work coming from calculating the $\frac{\del\varphi_\star}{\del\varphi_c}$, which we do by following the method of \cite{astro-ph/0603799}. The idea is to write $\frac{\del\varphi_\star}{\del\varphi_c}=\frac{\del\varphi_\star}{\del C}\frac{\del C}{\del\varphi_c}$, where $C$ are the $n_\varphi-1$ constants that define which trajectory we are on. Given that, we establish from (\ref{eq:thePathConstants})
\ba
\frac{\del\hat C^{\alpha,\beta}}{\del\varphi^\alpha_\star}&=&f_{(\alpha)}(\varphi^\alpha_\star),\\\nonumber
\frac{\del\hat C^{\alpha,\beta}}{\del\varphi^\beta_\star}&=&-f_{(\beta)}(\varphi^\beta_\star),
\ea
in which case we see 
\ba
\label{eq:dC_by_d_phi_star}
\frac{\del C^{\alpha}}{\del\varphi^\alpha_\star}&=&f_{(\alpha)}(\varphi^\alpha_\star),\\\nonumber
\frac{\del C^{\alpha}}{\del\varphi^{\alpha+1}_\star}&=&-f_{(\alpha+1)}(\varphi^{\alpha+1}_\star),
\ea
for $\alpha=1...n_\varphi-1$. Moreover, we have that (\ref{eq:thePathConstants}) yields
\ba
\label{eq:dCbydC}
\delta^{\alpha\beta}&=&\frac{dC^\alpha}{dC^\beta}=
   f_{(\alpha)}(\varphi^\alpha_c)\left.\frac{d\varphi^\alpha}{d C^{\beta}}\right|_c
  -f_{(\alpha+1)}(\varphi^{\alpha+1}_c)\left.\frac{d\varphi^{\alpha+1}}{d C^{\beta}}\right|_c,\quad\alpha,\;\beta=1...n_\varphi-1,
\ea
which, for a given $\beta$, gives $n_\varphi-1$ relations for the $\frac{d\varphi^\alpha}{dC^\beta}$. Now note that the final surface is defined to be a uniform energy density surface, and that from (\ref{eq:rhoBar}), (\ref{eq:cosmosusy}) energy density is given by
\ba
{\cal E}&=&\frac{1}{2}(\dot\varphi,\dot\varphi)+V=6\pi GW^2,
\ea
so the surface at $t_c$ is defined by
\ba
W_c&=&const,
\ea
which implies that
\ba
\left.\frac{d\varphi^\gamma}{d\hat C^{\beta}}\right|_c\left.\del_\gamma W\right|_c&=&0.
\ea
We may now combine this single relation, for a given $\beta$, with the $n_\varphi-1$ relations (\ref{eq:dCbydC}) to give the following evaluated at $t_c$.
\ba
\left(
\begin{array}{cccc}
\del_1W & \del_2 W & \del_3 W & ...\\
f_{(1)}     & -f_{(2)}     & 0               & ...\\
0             & f_{(2)}      & -f_{(3)}      & ...\\
\vdots
\end{array}
\right)
\left(
\begin{array}{c}
\frac{d\varphi^1}{dC^\beta}\\
\frac{d\varphi^2}{dC^\beta}\\
\frac{d\varphi^3}{dC^\beta}\\
\vdots
\end{array}
\right)
=
\left(
\begin{array}{c}
0  \\
\vdots
\\
1 \\
0 \\
\vdots
\end{array}
\right),
\ea
where the non-zero element of the column vector on the right-hand-side is at row $\beta+1$. These matrix equations are then solved to yield {\it exact} expressions for $\frac{d\varphi^\alpha_c}{dC^\beta}$, and we combine this with (\ref{eq:dC_by_d_phi_star}) allowing us to calculate
\ba
\frac{\del\varphi^\alpha_c}{\del\varphi^\gamma_\star}&=&\frac{d\varphi^\alpha_c}{dC^\beta}\frac{\del C^\beta}{\del\varphi^\gamma_\star}.
\ea
For the classes of superpotential identified earlier we find the following for the variation of $N$.
\begin{itemize}
\item
CLASS I. 
\ba
\frac{1}{4\pi G}\frac{\del N}{\del\varphi^\alpha_\star}&=&\frac{1}{m}\left\{\left.\frac{w_{(\alpha)}}{w_{(\alpha)}'}\right|_\star
      -\sum_\beta\left.\frac{w_{(\beta)}}{w_{(\beta)}'}\right|_c\frac{\del\varphi^\beta_c}{\del\varphi^\alpha_\star}\right\}
\ea
\item
CLASS II. 
\ba
\frac{1}{4\pi G}\frac{\del N^{(\beta)}}{\del\varphi^\alpha_\star}&=&\left.\frac{w_{(\beta)}}{w_{(\beta)}'}\right|_\star\delta^\beta_\alpha
      -\left.\frac{w_{(\beta)}}{w_{(\beta)}'}\right|_c\frac{\del\varphi^\beta_c}{\del\varphi^\alpha_\star}
\ea
\end{itemize}
The higher order derivatives of $N$ follow from these expressions by differentiation, and may be substituted into (\ref{eq:nzeta}) to find the spectral index, and (\ref{eq:fnl}) to find the non-Gaussianity parameter.

\section{Conclusions}
In this paper we have presented arguments in favour of using a perturbation variable for the scalar fields that transforms covariantly under field-space redefinitions, replacing the $\delta\varphi^\alpha$ that is currently used, and allowing one to more easily write expressions for physical quantities that are manifestly invariant under $\varphi^\alpha\rightarrow\varphi'^\alpha(\varphi)$. These variables resolve a number of inconsistent expressions that have appeared in the literature, and may be used to define a natural set of adiabatic and entropy perturbations along, and normal to, the background evolution of the scalar. In introducing this covariant picture we also see that a common definition of adiabatic and entropy perturbation are not defined covariantly, and so we identify definitions of such perturbations that are manifestly field-space covariant.

Thinking about the evolution of the scalar in terms of paths in field space also led to a natural definition of slow-roll parameters in terms of the Hamilton-Jacobi formulation, which itself is intimately connected to the notion of pseudo-supersymmetry. This pseudo-supersymmetry turned out to be a useful tool in calculating physical observables related to the perturbations of the scalars, allowing us to calculate expressions for the derivatives of $N$ (the number of e-folds) with respect to the scalars, and this is done without relying on the slow-roll approximation.

There are a number of avenues that remain to be explored. While we have presented two classes of superpotential that allow for explicit expressions of observables it is far from clear that this is the full set, it would be useful to know the general class of superpotential that can be solved analytically. An exploration of the physical implication of pseudo-susy is also required, and here the technique should be fruitful as we are able to have analytic control over the evolution of the fields for longer; pseudo-susy does not need the slow-roll approximation and so the evolution may be explored even after the end of inflation.

\acknowledgments
PS would like to thank Adam Christopherson, Tony Padilla, Costas Skordis and Ewan Tarrant for discussions, and STFC for financial support.

\appendix

\section{Scalar-field perturbations}
\label{app:scalarFieldPert}

Now we shall follow \cite{petrov,eisenhart} in setting up the Riemann co-ordinates for our problem, which specifies a co-ordinate system using geodesics. The first step is to solve the geodesic equation, which we do as a power series expansion in $\sigma$, the proper distance along a curve,
\ba
\varphi^\alpha(\sigma)&=&\varphi_{(0)}^\alpha+\sigma\left.\frac{d\varphi^\alpha}{d\sigma}\right|_0+\frac{1}{2}\sigma^2\left.\frac{d^2\varphi^\alpha}{d\sigma^2}\right|_0+...,
\ea
upon substitution of this into the geodesic equation (\ref{eq:geod}) we find
\ba
\label{eq:geodExp}
\varphi^\alpha(\sigma)&=&\varphi_{(0)}^\alpha+\sigma\xi^\alpha_{(0)}-\frac{1}{2}\sigma^2\Gamma^\alpha_{\;\beta\gamma}\xi^\beta_{(0)}\xi^\gamma_{(0)}
-\frac{1}{3!}\sigma^3\Gamma^\alpha_{\;\beta\gamma\delta}\xi^\beta_{(0)}\xi^\gamma_{(0)}\xi^\delta_{(0)}+...,
\ea
where
\ba
\Gamma^\alpha_{\;\beta\gamma\delta}&=&``\nabla_\delta"\Gamma^\alpha_{\;\beta\gamma},\;\Gamma^\alpha_{\;\beta\gamma\delta\epsilon}
=``\nabla_\epsilon"\Gamma^\alpha_{\;\beta\gamma\delta},...,
\ea
and $``\nabla_\alpha"$ is the "covariant" derivative that only sees the lower indices, e.g. $``\nabla_\delta"\Gamma^\alpha_{\;\beta\gamma}=\del_\delta\Gamma^\alpha_{\;\beta\gamma}-\Gamma^\epsilon_{\;\delta\beta}\Gamma^\alpha_{\;m\gamma}
-\Gamma^m_{\;\delta\gamma}\Gamma^\alpha_{\;\beta\epsilon}$.

Our new co-ordinates, $\tilde\varphi^\alpha$, are related to the original ones by (\ref{eq:geodExp})
\ba
\label{eq:phi_phiTilde}
\varphi^\alpha&=&\varphi_{(0)}^\alpha+\tilde\varphi^\alpha-\frac{1}{2}\Gamma^\alpha_{\;\beta\gamma}\tilde\varphi^\beta\tilde\varphi^\gamma+...,
\ea
yielding
\ba
\label{eq:jacobian}
\left.\frac{\del\varphi^\alpha}{\del\tilde\varphi^\beta}\right|_0&=&\delta^\alpha_\beta,
\ea
and so the co-ordinate transformation is invertible. We also note that the components of the tangent vector in these new co-ordinates are $\tilde\xi^\alpha=\frac{d\tilde\varphi^\alpha}{d\sigma}=\xi^\alpha_{(0)}$ so if we were to solve the geodesic equation as a series expansion for $\tilde\varphi^\alpha$ co-ordinates rather than $\varphi^\alpha$ we would find the analogue of (\ref{eq:geodExp}) to be
\ba
\label{eq:geodExpPhiTilde}
\tilde\varphi^\alpha(\sigma)&=&\sigma\xi^\alpha_{(0)}-\frac{1}{2}\sigma^2\tilde\Gamma^\alpha_{\;\beta\gamma}\xi^\beta_{(0)}\xi^\gamma_{(0)}+...,
\ea
where the $\tilde\Gamma^\alpha_{\;\beta\gamma}$ are the Christoffel symbols in the $\tilde\varphi$ co-ordinates. However, we know that the geodesics are given by $\tilde\varphi^\alpha(\sigma)=\sigma\xi^\alpha_{(0)}$, because that is what solve the geodesic equation (\ref{eq:phi_phiTilde}, \ref{eq:geodExp}), which tells us that $\tilde\Gamma^\alpha_{\;(\beta\gamma)}=\tilde\Gamma^\alpha_{\;(\beta\gamma\delta)}=...=0$. This reduces to 
\ba
\del_{(\alpha_1}\del_{\alpha_2}...\del_{\alpha_{n-2}}\tilde\Gamma^\beta_{\;\alpha_{n-1}\alpha_n)}&=&0,
\ea
which may be rewritten as
\ba
\del_\beta\del_{(\alpha_1}\del_{\alpha_2}...\tilde\Gamma^\gamma_{\;\alpha_{n-2}\alpha_{n-1})}
         &=&-\frac{2}{n-2}\del_{(\alpha_1}\del_{\alpha_2}...\alpha_{n-2}\tilde\Gamma^\gamma_{\;\alpha_{n-1})\beta}.
\ea
It is these explicit relations of the Christoffel symbols that make the perturbation analysis tractible, for example, in Riemann co-ordinates the Riemann curvature of the field-space is given by
\ba
\tilde R^\alpha_{\;\;\beta\gamma\delta}&=&\del_\gamma\tilde\Gamma^\alpha_{\;\beta\delta}-\del_\delta\tilde\Gamma^\alpha_{\;\beta\gamma},
\ea
which may be used to derive
\ba
\del_\gamma\tilde\Gamma^\alpha_{\;\beta\delta}&=&\frac{1}{3}\left[\tilde R^\alpha_{\;\beta\gamma\delta}+\tilde R^\alpha_{\;\delta\gamma\beta}\right],
\ea
One also finds that 
\ba
\del_{(\alpha_1}\del_{\alpha_2}\tilde\Gamma^\beta_{\;\alpha_3)\gamma}&=&-\frac{1}{2}D_{(\alpha_1} \tilde R^\beta_{\;\alpha_2|\gamma|\alpha_3)},\\
\del_{(\alpha_1}\del_{\alpha_2}\del_{\alpha_3}\tilde\Gamma^\beta_{\;\alpha_4)\gamma}
            &=&-\frac{3}{5}\left[ D_{(\alpha_1} D_{\alpha_2}\tilde R^\beta_{\;\alpha_3|\gamma|i_4)}
               +\frac{2}{9}\tilde R^\beta_{\;(\alpha_1\alpha_2|\delta|}\tilde R^\delta_{\;\alpha_3\alpha_4)\gamma}\right],
\ea
which corrects a typo in \cite{AlvarezGaume:1981hn}.

Having introduced our perturbation variable, we now need to expand the various quantites that appear in the scalar-field sector of the Lagrangian density (\ref{eq:scalarLagrangian}). To do this we note that the expansion of a general covariant tensor on the scalar manifold, in terms of our perturbation variable is \cite{AlvarezGaume:1981hn}
\ba
\label{eq:taylorExp}
T_{\alpha_1\alpha_2...\alpha_m}(\varphi_{(0)}+\delta\varphi)&=&
\sum_{n=0}^{\infty}\frac{1}{n!}\left[\frac{\del}{\del\tilde\varphi^{\beta_1}}...\frac{\del}{\del\tilde\varphi^{\beta_n}}T_{\alpha_1\alpha_2...\alpha_m}\right]_{0}
\chi^{\beta_1}...\chi^{\beta_n},
\ea
and we may use the properties of the Riemann co-ordinate system to derive, for example, 
\ba
\label{eq:ddT}
\del_{(\alpha_1}\del_{\alpha_2)}\tilde T_{\gamma\delta}&=&D_{(\alpha_1}D_{\alpha_2)}\tilde T_{\gamma\delta}
                               -\frac{1}{3}\left(\tilde R^\beta_{\;(\alpha_1|\gamma|\alpha_2)}\tilde T_{\beta\delta}
                               +\tilde R^\beta_{\;(\alpha_1|\delta|\alpha_2}\tilde T_{\gamma\beta}\right),\\\nonumber
\label{eq:dddT}
\del_{(\alpha_1}\del_{\alpha_2}\del_{\alpha_3)}\tilde T_{\gamma\delta}&=&D_{(\alpha_1}D_{\alpha_2}D_{\alpha_3)}\tilde T_{\gamma\delta}
         -\left(\tilde R^j_{\;(\alpha_1|\gamma|\alpha_2}D_{\alpha_3)}\tilde T_{\beta\delta}
         +\tilde R^\beta_{\;(\alpha_1|\delta|\alpha_2}D_{\alpha_3)}\tilde T_{\gamma\beta}\right)\\
  &~&-\frac{1}{2}\left(D_{(\alpha_1}\tilde R^\beta_{\;\alpha_2|\gamma|\alpha_3)}\tilde T_{\beta\delta}
     +D_{(\alpha_1}\tilde R^\beta_{\;\alpha_2|\delta|\alpha_3)}\tilde T_{\gamma\beta}\right),\\
\label{eq:ddddT}
\del_{(\alpha_1}\del_{\alpha_2}\del_{\alpha_3}\del_{\alpha_4)}\tilde T_{\gamma\delta}&=&
          D_{(\alpha_1}D_{\alpha_2}D_{\alpha_3}D_{\alpha_4)}\tilde T_{\gamma\delta}\\\nonumber
      &~&+2\left(\tilde R^\beta_{\;(\alpha_1\alpha_2|\gamma|}D_{\alpha_3}D_{\alpha_4}\tilde T_{\beta\delta}
         +\tilde R^\beta_{\;(\alpha_1\alpha_2|\delta|}D_{\alpha_3}D_{\alpha_4}\tilde T_{\gamma\beta}\right)\\\nonumber
      &~&-2\left(D_{(\alpha_1}\tilde R^\beta_{\;\alpha_2|\gamma|\alpha_3}D_{\alpha_4)}\tilde T_{\beta\delta}
         +D_{(\alpha_1}\tilde R^\beta_{\;\alpha_2|\delta|\alpha_3}D_{\alpha_4)}\tilde T_{\gamma\beta}\right)\\\nonumber
      &~&+\frac{3}{5}\left(D_{(\alpha_1}D_{\alpha_2}\tilde R^\beta_{\;\alpha_3\alpha_4)\gamma}\tilde T_{\beta\delta}
         +D_{(\alpha_1}D_{\alpha_2}\tilde R^\beta_{\;\alpha_3\alpha_4)\delta}\tilde T_{\gamma\beta}\right)\\\nonumber
      &~&+\frac{1}{5}\left(\tilde R^\beta_{\;\alpha_1\alpha_2|\epsilon|}\tilde R^\epsilon_{\;\alpha_3\alpha_4)\gamma}\tilde T_{\beta\delta}
         +\tilde R^\beta_{\;\alpha_1\alpha_2|\epsilon|}\tilde R^\epsilon_{\;\alpha_3\alpha_4)\delta}\tilde T_{\gamma\beta}\right)\\\nonumber
      &~&+\frac{1}{3}\tilde R^\epsilon_{\;(\alpha_1\alpha_2|\gamma|}\tilde R^\beta_{\;\alpha_3\alpha_4)\delta}\left(\tilde T_{\epsilon\beta}
                       +\tilde T_{\beta\epsilon}\right).
\ea
Now we notice that the right-hand-side of equations (\ref{eq:ddT}-\ref{eq:ddddT}) are composed of tensor quantities, and so when substituted into the Taylor expansion (\ref{eq:taylorExp}) we have a fully covariant expression form the terms of the perturbation expansion. These expressions allow us to compute $G_{\alpha\beta}(\varphi_{(0)}+\delta\varphi)$, and $V(\varphi_{(0)}+\delta\varphi)$, but we still need to find the expansion for $\del_\mu(\varphi_{(0)}+\delta\varphi)$ of the kinetic term. This is achieved by noting that the perturbation is at $\sigma=1$, and so (\ref{eq:chiDef}), (\ref{eq:geodExpPhiTilde}) give
\ba
\varphi^\beta_{(0)}+\delta\varphi^\beta=\varphi^\beta(\sigma=1)&=&\varphi^\beta_{(0)}+\chi^\beta
                          -\frac{1}{2}\left(\tilde\Gamma^\beta_{\alpha_1\alpha_2}\right)_0\chi^{\alpha_1}\chi^{\alpha_2}+...,\\\nonumber
\Rightarrow\del_\mu(\varphi^\beta_{(0)}+\delta\varphi^\beta)&=&\del_\mu\varphi^\beta_{(0)}
                            +\del_\mu\chi^\beta
    -\frac{1}{2}\left(\del_\gamma\tilde\Gamma^\beta_{\alpha_1\alpha_2}\right)_0\chi^{\alpha_1}\chi^{\alpha_2}\del_\mu\varphi^\gamma_{(0)}+...,
\ea
in Riemann co-ordinates, where we have used (\ref{eq:jacobian}). Now we use our relations for the Christoffel symbols in Riemann co-ordinates to show
\ba
\del_\gamma\tilde\Gamma^\beta_{\;(i_1i_2)}&=&-\frac{2}{3}\tilde R^\beta_{\;(i_1i_2)\gamma},\\
\del_\gamma\tilde\Gamma^\beta_{\;(i_1i_2i_3)}&=&\frac{1}{2}D_{(i_1}\tilde R^\beta_{\;i_2i_3)\gamma},\\
\del_\gamma\tilde\Gamma^\beta_{\;(\alpha_1\alpha_2\alpha_3\alpha_4)}&=&-4!\left(\frac{1}{60}D_{(\alpha_1}D_{\alpha_2}R^\beta_{\;\alpha_3\alpha_4)\gamma}
                              -\frac{1}{45}R^\beta_{\;(\alpha_1\alpha_2|\delta|}R^\delta_{\;\alpha_3\alpha_4)\gamma}\right),
\ea
leading to a covariant expression that does not rely on Riemann co-ordinates
\ba
\del_\mu(\varphi^\beta_{(0)}+\delta\varphi^\beta)&=&\del_\mu\varphi^\beta_{(0)}+D_\mu\chi^\beta
      +\frac{1}{3}R^\beta_{\;\alpha_1\alpha_2\gamma}\chi^{\alpha_1}\chi^{\alpha_2}\del_\mu\varphi^\gamma_{(0)}\\\nonumber
   &~&+\frac{1}{12}D_{\alpha_1}R^\beta_{\;\alpha_2\alpha_3\gamma}\chi^{\alpha_1}\chi^{\alpha_2}\chi^{\alpha_3}\del_\mu\varphi^\gamma_{(0)}\\\nonumber
   &~&+\left(\frac{1}{60}D_{\alpha_1}D_{\alpha_2}R^\beta_{\;\alpha_3\alpha_4\gamma}
      -\frac{1}{45}R^\beta_{\;\alpha_1\alpha_2\delta}R^\delta_{\;\alpha_3\alpha_4\gamma}\right)
                 \chi^{\alpha_1}\chi^{\alpha_2}\chi^{\alpha_3}\chi^{\alpha_4}\del_\mu\varphi^\gamma_{(0)}+...,
\ea
where we have introduced the covariant derivative
\ba
D_\mu\chi^\alpha=\del_\mu\chi^\alpha+\Gamma^\alpha_{\;\beta\gamma}\del_\mu\varphi^\beta\chi^\gamma.
\ea
In the main section of the paper we shall meet objects with spacetime indices, whos covariant derivatives are given in terms of the Christoffel symbols of the spacetime, $\left\{_{\mu\;\rho}^{\;\;\nu}\right\}$
\ba
\nabla_\mu X^\nu&=&\del_\mu X^\nu+\left\{_{\mu\;\rho}^{\;\;\nu}\right\}X^\rho,
\ea
we shall also have objects with both spacetime and scalar-field manifold indices, in which case the covariant derivatives are given by
\ba
\label{eq:spacetimeFieldCovDer}
{\cal D}_\mu\chi^\alpha=\del_\mu\chi^\alpha+\Gamma^\alpha_{\;\beta\gamma}\del_\mu\varphi^\beta\chi^\gamma+\left\{_{\mu\;\rho}^{\;\;\nu}\right\}X^\rho.
\ea
We also need the expansion of the potential, which turns out to be
\ba\nonumber
\label{eq:pertPot}
V(\varphi_{(0)}+\delta\varphi)&=&V(\varphi_{(0)})+\del_\alpha V_0\chi^\alpha+\frac{1}{2}D_{\alpha_1}\del_{\alpha_2}V_0\chi^{\alpha_1}\chi^{\alpha_2}
           +\frac{1}{3!}D_{\alpha_1}D_{\alpha_2}\del_{\alpha_3}V_0\chi^{\alpha_1}\chi^{\alpha_2}\chi^{\alpha_3}+...\\
\ea
\section{Cosmological perturbations}
\label{app:cosPerts}
Following the conventions in \cite{BROWN-HET-796}, except for a change of spacetime metric signature, the line element for the perturbed FRW cosmological spacetime and fluid four-velocity are 
\ba
\label{eq:perturbedLineElement}
ds^2&=&a^2(\eta)\left\{-(1+2\phi)d\eta^2+2\del_iB\;dx^id\eta+\left[(1-2\psi)\gamma_{ij}+2E_{|ij}\right]dx^idx^j\right\},\\
\label{eq:perturbedFluidVelocity}
U_\mu&=&a(\eta)\left(-1-\phi,\del_iv\right),
\ea
where $\gamma_{ij}$ are the components of the spatial metric, with curvature constant $K$; the verical bar in $E_{|ij}$ denotes the covariant derivative with respect to the spatial metric $\gamma_{ij}$; and we are considering only scalar perturbations. Writing the energy momentum tensor in the form
\ba
\label{eq:fluidEMtensor}
T^\mu_{\;\;\nu}&=&(\rho+P)U^\mu U_\nu+P\delta^\mu_\nu,
\ea
and using Einstein's equations
\ba
G_{\mu\nu}&=&8\pi G T_{\mu\nu},
\ea
allows us to derive the background relations
\ba
\label{eq:friedmann}
H^2&=&\frac{8\pi G}{3}\bar\rho-\frac{K}{a^2},\\
\label{eq:Hdot}
\dot H&=&-4\pi G(\bar\rho+\bar P)+\frac{K}{a^2},
\ea
and the perturbed relations
\ba
\label{eq:Hkappa}
H\kappa-\frac{\Delta\psi+3K\psi}{a^2}&=&-4\pi G\delta\rho,\\
\label{eq:kappaRel}
\kappa+\frac{\Delta\sigma_s+3K\sigma_s}{a^2}&=&-12\pi G(\bar\rho+\bar P)av,\\
\dot\sigma_s+H\sigma_s-\phi+\psi&=&0,
\ea
where we have introduced \cite{astro-ph/0009268}
\ba
\sigma_s&=&-a(B-E'),\\
\label{eq:kappaDef}
\kappa&=&\frac{1}{a}\left[3({\cal H}\phi+\psi')+\Delta(B-E')\right],\\
{\cal H}&=&a'/a,\;\;H=\dot a/a,\\
a(\eta)d\eta&=&dt,
\ea
and an overdot denotes differentiation with respect to $t$ ($dt=a(\eta)d\eta$), a prime corresponds to differentation with $\eta$, and $\Delta$ is the Laplacian associated to $\gamma_{ij}$.
\section{Gauge invariant variables}
\label{app:gaugeInvVar}
It is often convenient to use variables that are explicitly invariant under the choice of spacetime gauge, rather than making a specific gauge choice, so knowing how the various quantities transform is useful. Here we have that under a small co-ordinate transformation
\ba
\tilde x^\mu&=&x^\mu+\xi^\mu,
\ea
then
\ba
\tilde\phi&=&\phi-{\cal H}\xi^0-\xi'^0,\\
\tilde\psi&=&\psi+{\cal H}\xi^0,\\
\tilde \sigma_s&=&\sigma_s-a\xi^0,\\
\tilde\chi^\alpha&=&\chi^\alpha-\bar\varphi'^\alpha\xi^0,
\ea
in which case one finds the following gauge invariant variables
\ba
\label{eq:gaugeInvChi}
\chi^\alpha_{(\psi)}&=&\chi^\alpha-\frac{\varphi'^\alpha}{\cal H}\psi,\\
\psi_{(\sigma)}&=&\psi+\frac{\cal H}{a}\sigma_s.
\ea
\section{Adiabatic and entropy perturbations in the two-field case}
\label{app:twoField}
If we consider the two-field limit of (\ref{eq:deltaSigmaDef}) and (\ref{eq:deltaSDef}) (using $\delta\underline\varphi$ instead of $\underline\chi$ to match \cite{astro-ph/0009131}) we find
\ba
\delta\sigma&=&\frac{\dot\varphi^1}{\dot\sigma}\delta\varphi^1+\frac{\dot\varphi^2}{\dot\sigma}\delta\varphi^2\\
\delta S&=&-\frac{\dot\varphi^2}{\dot\sigma}\delta\varphi^1+\frac{\dot\varphi^1}{\dot\sigma}\delta\varphi^2\\
\ea
then define \cite{astro-ph/0009131}
\ba
\frac{\dot\varphi^1}{\dot\sigma}=\cos\theta,\qquad \frac{\dot\varphi^2}{\dot\sigma}&=&\sin\theta.
\ea
Now we try to construct a field redefinition from $(\varphi^1,\varphi^2)$ to $(\sigma,S)$. The infinitesimal limit of the above relations tell us that
\ba
\frac{\del\sigma}{\del\varphi^1}&=&\cos\theta,\;\frac{\del\sigma}{\del\varphi^2}=\sin\theta,\\
\frac{\del S}{\del\varphi^1}&=&-\sin\theta,\;\frac{\del S}{\del\varphi^2}=\cos\theta.
\ea
Now, for these to be a well-defined transformation we must have $\frac{\del^2\sigma}{\del\varphi^2\del\varphi^1}=\frac{\del^2\sigma}{\del\varphi^1\del\varphi^2}$ and $\frac{\del^2S}{\del\varphi^2\del\varphi^1}=\frac{\del^2S}{\del\varphi^1\del\varphi^2}$, implying that
\ba
-\tan\theta\frac{\del\theta}{\del\varphi^2}&=&\frac{\del\theta}{\del\varphi^1},\qquad\frac{\del\theta}{\del\varphi^2}=\tan\theta\frac{\del\theta}{\del\varphi^1},
\ea
which combine to show that $\theta=const$ is the only solution that leads to a well-defined field redefinition, i.e. if we want to use $(\sigma,S)$ as field variables then we can only do so if the background evolution is trivial, otherwise one will find incorrect evolution equations for the entropy and adiabatic perturbations. This is simply because the evolution equations require partial derivatives with respect to $(\sigma,S)$, but these variables cannot in general be used as co-ordinates, due to the above integrability conditions being violated, and so such derivatives are not well defined, meaning that the evolution equations for $\delta\sigma$ and $\delta S$ of \cite{astro-ph/0009131} are not correct.

\end{document}